\documentclass[utf8]{FrontiersinHarvard} 

\usepackage{url,hyperref,lineno,microtype,subcaption}
\usepackage[onehalfspacing]{setspace}
\usepackage[english]{babel}
\usepackage{draftwatermark}

\def\keyFont{\fontsize{8}{11}\helveticabold }

\def\firstAuthorLast{Chau {et~al.}} 
\def\Authors{J. L. Chau\,$^{1,*}$, A. Berera\,$^{2}$, D. Huyghebaert\,$^{1}$}

\def\Address{$^{1}$Leibniz Institute of Atmospheric Physics at the University of Rostock, K\"{u}hlungsborn, Germany \\
$^{2}$ School of Physics and Astronomy, University of Edinburgh, Edinburgh, EH9 3FD, United Kingdom
}
\def\corrAuthor{J. L. Chau}

\def\corrEmail{jchau@iap-kborn.de}

\SetWatermarkText{ }
\SetWatermarkAngle{45}
\SetWatermarkScale{1}
\SetWatermarkColor[gray]{0.85}

\begin{document}
\onecolumn
\firstpage{1}

\title[MLT Rogue Vertical Drafts]{Rogue Vertical Drafts in the Mesosphere and Lower Thermosphere: Evidence and Implications}
%
%
\author[\firstAuthorLast ]{\Authors} 
\address{} 
\correspondance{} 

\extraAuth{}

\maketitle

\begin{abstract}
Observational evidence of extreme vertical velocities ($|w| \ge 12.5$ m/s and at times greater than  50 m/s) in the mesosphere and lower thermosphere (MLT), has emerged in recent years. We refer to these events as Rogue Vertical Drafts (RVDs). They exceed five standard deviations of observed vertical velocities and appear as paired updraft–downdraft structures in varicose mode. Four-dimensional observations reveal that RVDs are intermittent, recurrent, and unpredictable. On average, they are expected to occur every $\sim$12 days during summer over Northern Norway, assuming a 1000 s interval. Different instruments may capture only portions of these events—for example, only upward or downward drafts when restricted to a single altitude range. Despite their rarity, their magnitudes and frequency suggest potential impacts on dust-sized matter escaping from planets, natural and anthropogenic space material, and MLT climate and processes. We propose that RVDs are a fundamental yet under-recognized feature of the MLT, underscoring the need for global observations to assess their prevalence and significance.

\keyFont{\section{Published: Front. Astron. Space Sci., doi:10.3389/fspas.2025.1716224, 2025.}}
 
\tiny
 \keyFont{ \section{Keywords: Extreme vertical velocities, Rogue vertical drafts, climate change monitoring, MLT observing techniques, space debris, noctilucent clouds, planetary life transfer} } 
\end{abstract}

\section{Introduction}

The mesosphere and lower thermosphere (MLT) have been called the "ignorosphere", mainly because the difficulty to observe them, when compared to their adjacent regions. Being the transition between the Earth's atmosphere and the near-space environment, the MLT host fascinating processes, such as the ablation of the great majority of extra terrestrial material entering the Earth's atmosphere. These material, in combination with the very cold temperatures over high latitudes during summer months, contribute to the occurrence of Noctilucent Clouds (NLC) and the associated radar and satellite versions, i.e., polar mesospheric summer echoes (PMSE) and polar mesospheric clouds, respectively \citep[e.g.,][]{rapp+lubken-2004,fritts+etal-2020}. 

In recent years, significant observational efforts, both from ground and from satellite platforms, have been made. Such observations have helped improve the understanding of global circulation dynamics, including the mean winds and planetary-scale waves. Most of the MLT studies have been focused on climatology or large-scale processes. Studies of higher spatio-temporal resolution processes have been limited to a few case studies, again due to the limited observing capabilities for the region. Nonetheless, one of the intriguing and recurrent features on regional and relatively high-resolution observations has been the occurrence of large vertical velocities.

Based on accepted residual circulation processes, the mean vertical velocities in the MLT are expected to be within the order of cm/s, and their variability in most general circulation models (GCMs) to be less than 1 m/s \citep[e.g.,][]{smith-2012}. Such variability has slightly increased with the advent of high-resolution GCMs, particularly in non-hydrostatic models \citep[e.g.,][]{Kunze+etal-2025}.

On the other hand, different sources of observations have shown that the standard variability of vertical velocities is within the order of a few m/s around 80-90 km and larger at high altitudes \citep[e.g.,][]{hoppe+fritts-1995b, gudadze+etal-2019}. On occasions, large vertical velocities ($|w| \ge 12.5$ m/s) have been observed with values much larger than five sigma variability \citep{chau+etal-2021b,hartisch+etal-2024}. In this work we call such events Rogue Vertical Drafts (RVDs). 

In the following sections we present observational and numerical modelling evidence of RVDs; discuss their implications to dust escaping planets, space material from meteors and satellites, and MLT climate and processes; discuss their recurrence based on observations and statistical mechanics arguments. Finally, we share our concluding perspectives on the topic.

\section{Evidence}

\subsection{Observations of vertical velocities in the MLT}
\label{sec:observations}

Measurements of vertical velocities in the MLT have primarily focused on their mean (synoptic) values, which are expected to be on the order of a few cm/s. Due to their relatively small magnitudes, direct measurements require high accuracy and appropriate spatial and temporal sampling. More recently, attention has shifted towards the measurement of instantaneous vertical velocities ($w'$), which are important for understanding MLT weather processes.

Mean vertical velocities in the MLT have been investigated using both radars and optical instruments. For radars, direct measurements have been conducted using polar mesospheric summer echoes (PMSE) as tracers \citep[][]{balsley+riddle-1984, hoppe+fritts-1995} and incoherent scatter measurements \citep{zhou-2000, Oyama+etal-2005}, while indirect measurements have utilized specular meteor radars and partial reflection radars. Initial direct measurements of mean vertical winds showed discrepancies with the expected residual circulation, revealing larger velocities and opposite directions to the anticipated summer downward motions of a few cm/s \citep[e.g.,][]{hoppe+fritts-1995}. \cite{gudadze+etal-2019} demonstrated that PMSE-based mean vertical velocity estimates are influenced by the sedimentation of ice particles and, more importantly, by the spatio-temporal characteristics of PMSE, which prevent uniform sampling of both positive and negative velocities.

Indirect radar measurements of mean vertical velocities have either incorporated residual circulation to infer vertical velocities from meridional winds at different polar latitudes \citep[e.g.,][]{vincent+etal-2019}, or applied the continuity equation using precise gradients of horizontal winds \citep[e.g.,][]{laskar+etal-2017,zheng+etal-2024}. In both cases, reasonable small downward mean velocities have been obtained.

\begin{figure}[h!]
\begin{center}
\includegraphics[width=\textwidth]{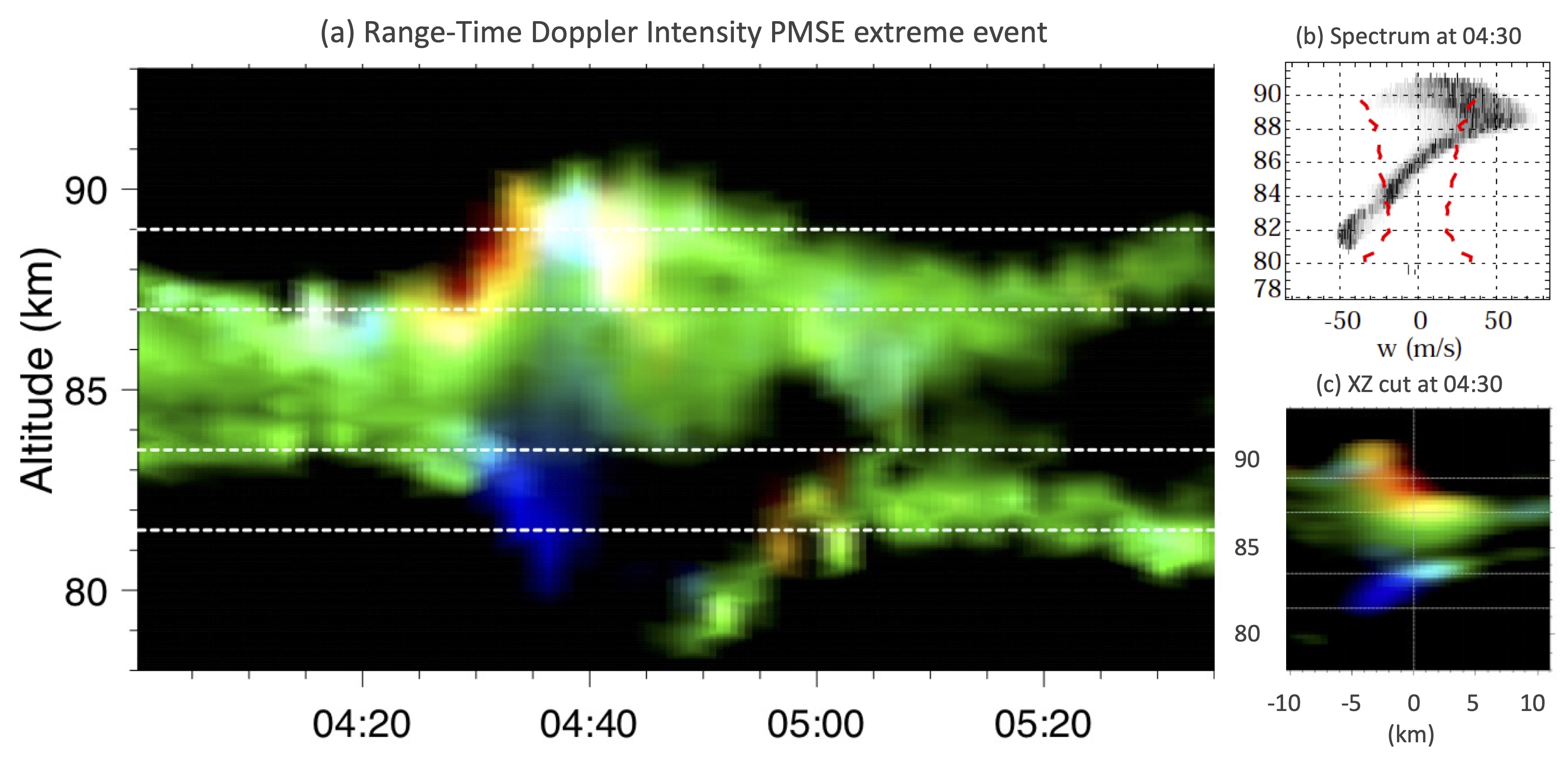}
\end{center}
\caption{An example of an RVD observed over northern Norway on July 16, 2016. (a) A Altitude-Time-Doppler Intensity plot of PMSE, color-coded with vertical velocities. (b) Spectrogram at 0430 UT showing extreme updraft and downdraft regions. The dashed red lines indicate the 3-sigma values during that season. (c) A horizontal-altitude cut around 0430 UT of PMSE, also color-coded with vertical velocities. Adapted from \cite{chau+etal-2021b}.}
\label{fig:rvd_maarsy}
\end{figure}

For $w'$, both radars and resonance lidars have been used \citep[e.g.,][]{gardner+liu-2007}. Lidar measurements over short periods and small volumes reveal $w'$ values of a few m/s, occasionally lasting tens of minutes, and displaying both positive and negative excursions \citep[e.g.,][Figures 1c and 1d]{chu+etal-2022}. PMSE-based and ISR D-region measurements also show $w'$ variability of a few m/s \citep[e.g.,][]{hoppe+fritts-1995, zhou-2000, gong+etal-2017, gudadze+etal-2019}. Nonetheless, \cite{chau+etal-2021b} reported an extreme event characterized by updrafts reaching 60 m/s in the upper PMSE region, accompanied by downdrafts of -50 m/s in the lower region, both lasting several minutes over the relatively small volume of the radar. These magnitudes were over five times the typical $w'$ standard deviation (see Figure \ref{fig:rvd_maarsy}). More recently, \cite{hartisch+etal-2024} reported additional events over northern Norway, showing updrafts and downdrafts lasting several minutes, with velocities up to 3–4 times the standard deviation, i.e., less than 30 m/s.

The 557.7~nm optical emissions from the lower thermosphere have also provided opportunities to monitor vertical winds \citep{larsen+meriwether-2012}.  By using a Fabry-Perot interferometer it is possible to obtain the vertical wind emissions by measuring the Doppler shift of the emissions.  Instances of vertical winds greater than 50~m/s have been detected previously with these measurements, though one potential issue with the 557.7~nm emission is that the determination of the altitude of the origin of the emissions can be variable throughout a measurement period during active auroral conditions \citep{larsen+meriwether-2012}.  Multiple instances of vertical winds greater than 20 m/s lasting several hours existed in the different datasets presented.

Airglow measurements from hydroxyl (OH) in the MLT have also been used to infer vertical wind velocities at polar latitudes in the winter using a Michelson Interferometer \citep{bhattacharya+gerrard-2010}.  The emission rate of OH peaks at approximately 84~km in altitude.  Data from multiple arctic winter campaigns were analysed to determine the vertical winds at the mesopause altitude.  Velocities within $\pm$~20~m/s were common, with some events reaching speeds greater than this.  A potential link of the daily variance in vertical winds to the polar vortex location was suggested.

These observations remain limited to specific geographic locations and times of year. For instance, resonance lidar observations are constrained to clear nighttime conditions, while PMSE observations are limited to the summer polar season when echoes are present. To address these sampling limitations, efforts have been made to retrieve MLT vertical velocities—both mean and perturbations—using specular meteor radars (SMRs), which operate continuously regardless of weather conditions and at all latitudes.

Most SMRs operate in monostatic mode and traditionally provide horizontal velocities averaged over a volume with a 200 km radius, 2 km vertical resolution, and 1–2 hour intervals, assuming zero mean vertical velocity. Occasionally, estimates of mean vertical velocities have been attempted, yielding values significantly larger than expected from tidal theory \citep[e.g.,][]{egito+etal-2016}. \cite{chau+etal-2017} noted that monostatic systems can yield biased vertical velocity estimates due to horizontal velocity gradients, necessitating corrections that require multistatic SMRs (MSMRs). Despite these corrections, MSMR-based vertical velocity estimates remain large (10–15 m/s) and persist for several hours \citep[e.g.,][]{chau+etal-2021, conte+etal-2021, charuvil+etal-2022}. \cite{charuvil+etal-2022a}, using a virtual MSMR setup within a regional atmospheric model, demonstrated that small-scale variability in horizontal winds, if not uniformly sampled, contributes to biases in vertical velocity estimates. Additionally, uncertainties in measurement positions, the nature of the scattering region, and the precision of one-dimensional velocity projections also affect these estimates \citep[e.g.,][]{stober+etal-2022}.

To address these challenges while still pursuing vertical velocity retrievals, \cite{urco+etal-2024} implemented a physics-informed neural network (PINN) approach called HYPER (HYdrodynamic Point‐wise Environment Reconstructor), which integrates the Navier–Stokes equations into the velocity inversion process. HYPER provides $w'$ estimates over a 50 km $\times$ 50 km $\times$ 1 km domain every 30 minutes, with magnitudes of a few m/s, while mean values remain on the order of a few cm/s. These $w'$ values are consistent with high-resolution non-hydrostatic atmospheric model simulations. However, RVD events have not yet been reported from HYPER analyses.

\subsection{Sources for creating vertical winds}
The RVDs mentioned above are consistent with predictions from direct numerical simulations (DNS) of stratified flows. \cite{feraco+etal-2018} predicted intense, localized vertical velocities in space and time under specific stratification conditions, particularly for Froude numbers in the range $\sim$0.1–0.01. Amplification of $w'$ has also been obtained in 2D DNS simulations of mesospheric bores, where ducted atmospheric regions were perturbed by waves, with greater amplification for thinner ducts \citep[e.g.,][]{ramachandran+etal-2023}. Using the CGCAM (Complex Geometry Compressible Atmospheric Model), a finite-volume code solving the compressible Navier–Stokes equations, \cite{lund+etal-2020} simulated large MLT vertical velocities of several tens of m/s over mountainous regions.

\section{Implications}

\subsection{Space dust planetary escape mechanism}

Observations find that hypervelocity space dust at speeds
$\approx 10-70~{{\rm km}/s}$ continuously bombard the Earth, at the level of
$\approx 10^5$ kilograms per day
\citep{kw86,lb93,flynn2002,csfbkch,plane}.
Such fast space dust produces immense
momentum flows in the higher atmosphere from the upper mesosphere and above.  In \cite{absd} it was pointed out that the speed of this space dust
is sufficiently large that in collision with particles in the
higher atmosphere, it could transfer considerable momentum,
enough for some particles to reach escape velocity and
leave Earth's gravitational pull.  Such particles could
be constituents that comprise the upper atmosphere, so
the various atoms and molecules that form it, but these
particles might also be tiny pieces of biological material, such
as DNA or microbial life, and propel them into space free of
Earth.  The latter possibility is particularly interesting
to origin of life questions.  One common idea is that life
began on Earth and possibly other planets by initially
entering the planet from outer space.  This space
dust planetary escape mechanism would provide one means
for biological particles to leave the gravitational
pull of its host planet.

Estimates on the flux of space dust bombarding Earth based on ground
and satellite measurements show there are approximately
$10^9$ grams of space dust per year for each decade of particle
mass from space dust grain masses ranging from $10^{-9}~{\rm g}$
to $10^{-2}~{\rm g}$ \citep[][]{kw86,lb93,flynn2002,csfbkch,plane}.
This space dust will enter Earth's atmosphere moving in
all different directions. A fraction of this space dust will
have a grazing trajectory, so enter and then exist Earth's atmosphere.
At $150$ km,
\cite{absd} estimated this to be around $20 \%$ of the total
space dust flux.  A typical scale for a small biological
particle is a radius at or smaller than $10^{-6}$ m
with a typical matter density
of $1-2 \times 10^3~{\rm kg/m^3}$, so giving a mass $\sim10^{-11}~{\rm g}$.
This is much less massive than a major portion of space dust
grains, so collisions would accelerate such biological
particles to the speed of the colliding fast space dust particle,
thus often to above escape velocity.
However friction with
the atmosphere could drag these particles and heat them up.
Only above around $150$ km it was found in \cite{absd} that the
atmospheric density is low enough for atmospheric drag
and heating to be negligible for small particles at escape
velocity speed.

At an altitude of $150$ km biological material is not expected to
be commonly found.  At sea level and into the troposphere
and even up to the middle of the stratosphere
there are small particles up to micron size
including biological material found
\citep[][]{rosen64,xszzi,hht,ycjczz,ursem16}.
At higher altitude noctilucent clouds are present $ 80-90 $ km above sea level, so if water vapour can reach that high up one could expect perhaps there is also biological material present. However above this altitude one would
not expect any large quantity of biological material
reaching by natural means.
Thus for this escape mechanism to be effective, it requires that these small biological organisms be transported from the troposphere, where they are abundantly present, up to the lower thermosphere. There are various mechanisms that provide upward forces on small particles in the atmosphere such as hurricanes and extreme events in the lower atmosphere and volcanoes but these can only push up particles into the stratosphere, perhaps at best into the lower mesosphere \citep[][]{verbeek1884,sr81,ludlam57,wshw78,tthgr09}.
Gravito-photophoresis arising from sunlight irradiating particles can elevate micron scale particles up to $80$ km \citep[][]{rohatschek96}.
However above this altitude mesospheric and thermospheric vertical winds would be the most likely means \citep[][]{absd,abdbvw}.

To calculate the effect of vertical winds on altitude
climb for a test particle, \cite{abdbvw} developed a simple model. This model recognizes a symmetry that
the profile of vertical winds is similar over different horizontal
positions over the Earth, thus allowing the complicated
horizontal dynamics to be factored out at zeroth
approximation. It is then a one-dimensional problem
to estimate the particle altitude climb.  Vertical
winds also vary in velocity over time, but to simplify
they consider a constant upward vertical wind.
This captures the basic physics and allows for simple
estimates.  \cite{abdbvw} considered the climb for a disc shaped particle of
density $\rho_p$, radius $r$ and height $h$, thus the mass
being $m_p = \pi r^2 h \rho_p$.
The equation obtained from this model for the
evolution of a test particle from a vertical wind is,
\begin{equation}
\frac{dv}{dt} = -g + \frac{\rho(z)}{\rho_p h} (w(z,t) - v(t))^2 \;,
\end{equation}
where $v$ is the upward speed of the particle (negative being downward), $g \approx 9.8~{\rm m/s^2}$ is the gravitational acceletation near Earth, $w>0$ is the upward speed of the vertical wind, $\rho(z)$ the density of the atmosphere at the specified altitude $z$, and $h$ the height of the test particle.  This equation is valid for $w > v(t)$. When $dv/dt=0$, this gives the steady state velocity of the test particle of $v(t) = w - v_{th}$, where $v_{th} = \sqrt{g \rho_p h/\rho(z)}$ is the threshold velocity, which is the minimum upward velocity the vertical wind must have for the test particle to move upward. At each altitude $z$, $v_{th}$ depends only on the properties of the test particle and the atmospheric density. For example at altitude $z=100~{\rm km}$, $\rho(z) = 5.604 \times 10^{-7}~{\rm kg/m^3}$ (U.S. Standard Atmosphere value). Biological material have internal density ranging $1-2 \times 10^3~{\rm kg/m^3}$ and size as small as some tenths of a ${\rm nm}$ for DNA molecules to tens of ${\rm nm}$ for viruses or nanobes to hundreds of ${\rm nm}$ for small bacteria. This gives $v_{th({\rm DNA})} \approx 3~{\rm m/s}$, $v_{th({\rm small \ virus})} \approx v_{th({\rm nanobe})} \approx 22~{\rm m/s}$, and 
$v_{th({\rm small \ bacteria})}\approx 42~{\rm m/s}$.
Thus for Rogue Vertical Drafts reported in the previous section of $\sim 10~{\rm m/s}$ for minutes to hours,
these would drive up DNA molecules, whereas updrafts
reported up to $60~{\rm m/s}$ could
push up nanobe, small viruses, and small bacteria.  
                  
In summary, large upward vertical winds in the MLT at speeds above $\sim 10\ {\rm m/s}$, like those in RVDs, can provide sufficient force for the size and internal density of small biological material to push them upward in the atmosphere, against the force of gravity. For such particles to attain large climbs by this process one possibility is from large sustained upward drafts.  Another possibility is in general the vertical winds will have both upward and downward drafts as function both of time at a given location or at different locations. Also such particles will be subjected to  the effect of large horizontal winds.  This can create a random walk type motion of the test particles,  therefore some proportion of such particles will reach the extremes of the distribution and attain sizeable climbs in altitude.

\subsection{Climate science}

The occurrence and duration of extreme weather events have been shown to be increasing in the Troposphere with links to climate change \cite[e.g.,][]{ummenhofer+meehl-2017}.  It can be expected that there will also be associated increases in extreme weather events at higher altitudes due to interactions between the different atmospheric regions.  There are links between Sudden Stratospheric Warmings (SSW) and the wave-influenced coupling of the stratosphere-mesosphere-thermosphere system, with significant effects on the temperature and circulation of the mesosphere during these events \citep{chandran+etal-2014}.  SSW have also been shown to be increasing in length from the 1980s to the 2010s by 50\%, from 10 days to 15 days, with a potential attribution to climate change \citep{li+etal-2023}.  These SSW cause strong perturbations from the mean circulation patterns of the mesosphere, which increases the probability of RVDs occurring.

\subsection{Natural and anthropogenic space material}
The understanding of extreme MLT weather is important due to the different complex chemical processes that occur related to the heavy metal ions in the region.  The transport of metallic species from ablated meteoric material can have important implications for different reactions, including those involving ozone \citep{plane+etal-2015}.  Vertical winds and the transport of heavy metal ions from 90 to 100 km can also increase the lifetimes of these ions by orders of magnitude.  The metals can further affect sporadic E-layer formation and radio wave propagation in the region.  By monitoring MLT RVDs consistently over long periods of time, one can better understand how the changing terrestrial climate is influencing the upper atmospheric layers.

At altitudes below approximately 85~km space debris begins to ablate in the terrestrial atmosphere.  This releases exotic materials into the region which can have significant impact on the atmospheric chemistry.  Upward vertical drafts can move this ablated material to higher altitudes, which can result in significantly longer lifetimes before molecularization of the metallic atoms below 80~km and ultimately sedimentation \citep{plane+etal-2015}.  In the opposite case, downward RVDs could quickly reduce the atomic metal content in the MLT through the transport of recently ablated material to regions where the metallic atoms would be converted to molecular species through chemical processes.

\section{Theory - justification for presence of fast vertical winds}
\label{sec:theory}

We interpret Rogue Vertical Drafts as extreme weather
events in the higher atmosphere, the mesosphere and
above.  Where the hydrostatic approximation is a 
successful mean description of the atmosphere,
RVDs should be understood as extreme and relatively infrequent fluctuations about this mean.  This interpretation
thus implies a probability distribution 
function (PDF) that
quantifies vertical wind behavior, with RVDs at the 
outer tails of such a distribution.  Here we
construct one such example of a PDF from vertical wind data measured by the Middle Atmosphere Alomar Radar System (MAARSY) radar \citep{Latteck:2012}.

\begin{figure}[h!]
\begin{center}
\includegraphics[width=\textwidth]{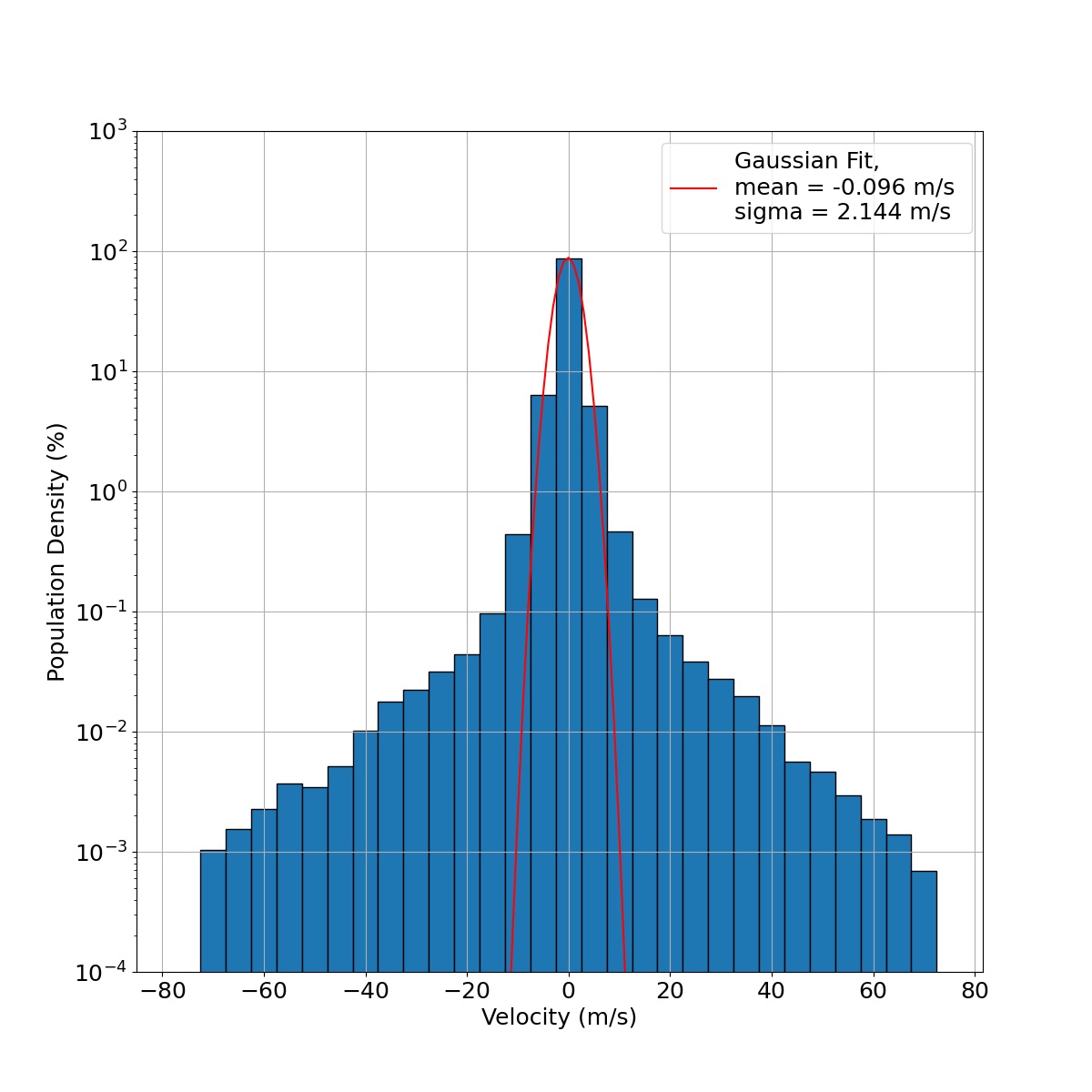}
\end{center}
\caption{Distribution of vertical velocities measured by MAARSY in Northern Norway.  Adapted from \cite{hartisch+etal-2024}.}
\label{fig:vertical_wind_dist}
\end{figure}

The total vertical wind statistics reported by \cite{hartisch+etal-2024} are displayed in Figure~\ref{fig:vertical_wind_dist} as a population density in percentage.  Each of these measurements were from PMSE with MAARSY and are therefore from altitudes between 80-90~km in the summer months (June-August).  The time and range resolutions are approximately 100~s and 300~m.  Typically PMSEs will cover 1-10~km in extent, and the winds can be relatively consistent for 10s of minutes.  If we consider that the events last $\approx$ 1000~s (10 measurement time intervals), at an occurrence of 1/1000 for events with updrafts greater than 12.5~m/s, the updraft RVDs could occur every $\approx$ 12 days based on the occurrence rates in Figure~\ref{fig:vertical_wind_dist}.  These estimates focus on the summer months, therefore different measurement techniques for other times of the year would be needed - such as meteor trail derived vertical winds, airglow measurements, or LIDAR.

A Gaussian fit is made to the original velocity data, with the resulting population density of the fit shown by the red line in Figure~\ref{fig:vertical_wind_dist}.  The function that is fit is:
\begin{equation}
f(x) = A \exp{\bigg[-\frac{(x-\mu)^2}{2{\sigma}^2}\bigg]}
\end{equation}
Where A is the amplitude, $\mu$ is the mean, and $\sigma$ is the standard deviation.  With this data, a 5$\sigma$ difference is $\approx$ 10.72~m/s.  The standard deviation of the dataset without fitting a normal distribution is 2.624 m/s.  12.5~m/s would be considered a sufficiently extreme event if the data is expected to follow a normal distribution.  To summarize, we have an expected rate of approximately one 15~minute vertical draft event with a velocity greater than 12.5~m/s every 12-days.  Note again that this is for data in Northern Norway during the summer using PMSEs as a tracer and the data are therefore limited to altitudes of 80-90~km.  The RVD occurrence rate could vary greatly by altitude, season, and measurement location.  Tornados and hurricanes, as examples, do not commonly occur in all regions of the globe.

Summertime incoherent scatter from the Troms\o~UHF radar were also previously investigated at altitudes of 96-111~km to examine the vertical wind velocities in the MLT region \citep{Oyama+etal-2005}.  Distributions showed velocities of $\pm$~20~m/s commonly, with extreme cases of speeds greater than this.  The PDF for these measurements had standard deviations of less than 15~m/s.  This would mean an increase in the standard deviation with altitude if we consider Figure~\ref{fig:vertical_wind_dist} and the resulting standard deviation found, corresponding to an increase in the likelihood of large vertical drafts at higher altitudes.  This is not unexpected due to the decreasing atmospheric density with altitude.


\section{Conclusions}

Rogue Vertical Drafts represent a unique and identifiable extreme phenomenon in the mesosphere and lower thermosphere (MLT). We have reviewed observational evidence for these events and offered perspectives on their broader significance, extending beyond atmospheric science to astrobiology, climate processes, and the dispersal of natural and anthropogenic space material.

Future research should prioritize determining the spatial and temporal extent of RVDs and assessing whether their global occurrence rates depend on local geophysical conditions. Equally critical is uncovering the physical mechanisms driving these extreme drafts. Progress on these fronts will require enhanced instrument duty cycles, a broader global network of high-resolution MLT wind measurements, and comprehensive theoretical and numerical modeling once a sufficiently robust observational database is available.

\section*{Conflict of Interest Statement}

The authors declare that the research was conducted in the absence of any commercial or financial relationships that could be construed as a potential conflict of interest.

\section*{Author Contributions}
All authors have contribute to writing and revision. In addition, DH generated Figure \ref{fig:vertical_wind_dist}.


\section*{Funding}
This work was partially supported by the HONDA project funded by the German Science Foundation (DFG 552554225).

\section*{Acknowledgments}
The authors thank Jennifer Hartisch for providing the vertical wind statistics used in Figure \ref{fig:vertical_wind_dist}.


\section*{Data Availability Statement}
The data used in the figure of this paper have been already published (https://doi.org/10.22000/1688). References to those publications are included.

\bibliographystyle{Frontiers-Harvard} 
\bibliography{vertical}





\end{document}